\documentclass[a4paper,10pt]{article}

\usepackage{amsmath}
\usepackage{amssymb}
\usepackage{amsfonts}
\usepackage{amscd}
\usepackage{latexsym}
\usepackage{amsbsy}
\usepackage{bbm}
\usepackage[english]{babel}
\usepackage{graphicx}
\usepackage{psfrag}

\begin{document}

 \begin{center}

 \vspace{.5cm}

 {\Large {\bf Asymptotically Schr\"odinger Space-Times: }}\\[.5cm]
 {\Large {\bf TsT Transformations and Thermodynamics}}

 \vspace{.7cm}

 \begin{center}

 {\bf Jelle Hartong and  Blaise Rollier}\\[0.2cm]

 \vskip 25pt

 {\em  Albert Einstein Center for Fundamental Physics,\\
 Institute for Theoretical Physics,\\
 University of Bern,\\
 Sidlerstrasse 5, CH-3012 Bern,
 Switzerland\vskip 10pt}

 {email: {{\tt hartong@itp.unibe.ch}} and {\tt rollier@itp.unibe.ch}}
 \vskip 15pt

 \end{center}

 \vspace{1cm}

 {\bf Abstract}

 \begin{quotation}

 {\small
\noindent We study the complete class of 5-dimensional asymptotically Schr\"odinger space-times that can be obtained as the TsT transform of an asymptotically AdS$_5$ space-time. Based on this we identify a conformal class of Schr\"odinger boundaries. We use a Fefferman--Graham type expansion to study the on-shell action for this class of asymptotically Schr\"odinger space-times and we show that its value is TsT invariant. In the second part we focus on black hole space-times and prove that black hole thermodynamics is also TsT invariant. We use this knowledge to argue that thermal global Schr\"odinger space-time at finite chemical potential undergoes a Hawking--Page type phase transition.}

 \end{quotation}

 \end{center}

 \newpage

\tableofcontents

\section{Introduction}

Asymptotically Schr\"odinger space-times form a natural starting point to try to extend holographic techniques to the realm of nonrelativistic physics. There exists a nonrelativistic counterpart of a CFT that is based on the Schr\"odinger group \cite{Nishida:2007pj} and from a phenomenological point of view they could be of relevance to the description of cold atoms at unitarity \cite{Son:2008ye,Balasubramanian:2008dm}. The Schr\"odinger space-time can be thought of as a deformation of an AdS space-time. This deformation is such that certain properties of the AdS space-time carry over to the Schr\"odinger case. From the point of view of constructing holographic techniques the interpretation that a Schr\"odinger space-time is a deformation of an AdS space-time is both reassuring that this may be achievable as well as a useful guideline. On the other hand it is from the point of view of describing interesting nonrelativistic phenomena a bit of a nuisance because one needs to make sure that the phenomenon crucially depends on the deformation and not on the AdS part of the Schr\"odinger space-time. For example the properties of the propagators for a free scalar field theory can be inferred from AdS propagators \cite{Volovich:2009yh,Barnes:2010ev,Fuertes:2009ex}. As another example of physics that is insensitive to the deformation we will show that a class of asymptotically Schr\"odinger black holes has the same thermodynamics as the undeformed asymptotically AdS black hole.

What underlies the relation between asymptotically Schr\"odinger space-times (ASch) and asymptotically AdS (AAdS) space-times is a transformation known interchangeably as a TsT (T-duality, shift, T-duality) transformation \cite{Lunin:2005jy} or a Melvin twist \cite{Gimon:2003xk}. We will refer to it as a TsT transformation. Such a transformation can be thought of as a solution generating transformation relating different solutions of type II supergravities. In order to perform a TsT transformation one needs two compact commuting Killing vectors that form the isometries of a 2-torus. Suppose that the two torus directions are parametrized by $\xi$ and $V$. The transformation then involves a T-duality along $\xi$, followed by a shift $V\rightarrow V+\gamma\tilde\xi$ where $\tilde\xi$ parametrizes the T-dual cirlce and a second T-duality along $\tilde\xi$. If the initial field configuration was a solution of type IIA/B supergravity then the final field configuration will be another solution of type IIA/B supergravity. When this solution generating technique is applied to space-times of the form AAdS$_5\times S^5$ with $\xi$ parametrizing a circle in the 5-sphere and $V$ parametrizing a compact Killing vector of the AAdS$_5$ space-time that asymptotically becomes null then the resulting space-time is asymptotically Schr\"odinger (with a compact circle that becomes asymptotically null) and the TsT transformation is referred to as a null dipole deformation \cite{Maldacena:2008wh}. From now on when we talk about a TsT transformation we will mean a null dipole deformation. On the level of writing down asymptotically Schr\"odinger solutions the fact that the $V$ coordinate was assumed to be compact plays no role, the TsT transformed AAdS$_5$ space-times also exist when we take $V$ to be non-compact.

Up to date the only asymptotically Schr\"odinger space-time that was explicitly studied is the TsT transformed black brane solution of \cite{Maldacena:2008wh,Herzog:2008wg,Adams:2008wt,Imeroni:2009cs}. We will employ the TsT transformation to study a large class of ASch$_5$ space-times and characterize properties that are inherited from the AAdS$_5$ space-time. A dominant role in our analysis is played by the Killing vector $N$ of the AAdS$_5$ space-time that becomes asymptotically null and that is used in the TsT transformation. We will focus on those AAdS$_5$ space-times that admit $N$ as boundary null Killing vector and study the associated conformal class of boundaries and interpret their images under TsT as providing a conformal class of Schr\"odinger boundaries. This allows us to study counterterms for this class of Schr\"odinger boundaries. The main result of this analysis will be that the on-shell action is TsT invariant. We then continue to study thermodynamic properties of TsT transformed AAdS$_5$ black holes and show that also thermodynamic quantities such as entropy, temperature and chemical potentials are all TsT invariant. We end with a discussion of the Schr\"odinger analogue of the Hawking--Page type phase transition.

\section{Fefferman--Graham coordinates and asymptotically
Schr\"odinger space-times}\label{sec:FGandTsT}

The details of the TsT transformation are summarized in appendix \ref{app:TsT}. The result that will be most important to us is the form of the 5-dimensional fields $A_\mu$, $\phi$ and $\bar g_{\mu\nu}$ that result from applying a TsT transformation to a pure AAdS$_5$ space-time, i.e. without matter sources, which for the convenience of the reader are copied from appendix \ref{app:TsT} and repeated here
\begin{eqnarray}
 A & = & \gamma e^{2\Phi}g_{V\mu}dx^\mu\,,\label{eq:TsTvectorbody}\\
e^{-2\Phi} & = & 1+\gamma^2g_{VV}\,,\label{eq:TsTscalarbody}\\
\bar g_{\mu\nu} & = & e^{-2\Phi/3}\left(g_{\mu\nu}-e^{-2\Phi}A_\mu A_\nu\right)\,,\label{eq:TsTmetricbody}
\end{eqnarray}
where $g_{\mu\nu}$ is an AAdS$_5$ metric that has $\partial_V$ as a Killing vector. As shown in appendix \ref{app:TsT} the isometries of the TsT transformed metric $\bar g_{\mu\nu}$ are precisely all the AAdS$_5$ Killing vectors that commute with the AAdS$_5$ Killing vector $\partial_V$.

The class of ASch$_5$ space-times that we will focus on are all space-times that result from applying TsT to those pure AAdS$_5$ space-times that admit a Killing vector $N=\partial_V$ that asymptotically commutes with the Schr\"odinger algebra, so that $N$ is asymptotically null. This follows from the fact that the embedding of the Schr\"odinger algebra into the AdS isometry algebra is essentially unique \cite{Blau:2010fh} and that this embedding is such that the Schr\"odinger algebra consists of all those AdS$_5$ isometries that commute with a null Killing vector.

We emphasize that not every TsT transformation of a pure AAdS$_5$ space-time gives rise to an ASch$_5$ space-time. For example the stationary black hole solution (with spherical horizon topology) of \cite{Yamada:2008if} can be obtained as a TsT transformation of Schwarzschild-AdS$_5$ using a Killing vector that does not asymptotically commute with the Schr\"odinger algebra. Therefore the black hole solution of \cite{Yamada:2008if} is not asymptotically Schr\"odinger. This can also be inferred from the fact that the curvature invariants of this black hole solution do not asymptote to their Schr\"odinger values.

We will next introduce Fefferman--Graham (FG) coordinates \cite{FeffermanGraham} and discuss the properties of the class of pure AAdS$_5$ metrics that give rise to ASch$_5$ space-times. In FG coordinates the pure AAdS$_5$ metric $g_{\mu\nu}$ is given by
\begin{equation}
ds^2=g_{\mu\nu}dx^\mu dx^\nu=\frac{dz^2}{z^2}+h_{ab}(z,x)dx^adx^b\,,
\end{equation}
where 
\begin{equation}
h_{ab}(z,x)=\frac{1}{z^2}\left(g_{(0)ab}+z^2g_{(2)ab}+z^4g_{(4)ab}+\mathcal{O}(z^5)\right)\,,
\end{equation}
in which $g_{(2)ab}$ and $g_{(4)ab}$ are given by \cite{de Haro:2000xn}
\begin{eqnarray}
 g_{(2)ab} & = & -\frac{1}{2}\left(R_{(0)ab}-\frac{1}{6}R_{(0)}g_{(0)ab}\right)\,,\label{eq:g2intermsofg0}\\
 g_{(4)ab} & = & \frac{1}{4}a_{(4)}g_{(0)ab}+\frac{1}{2}g_{(2)ac}g_{(2)b}^c+\frac{1}{24}R_{(0)}g_{(2)ab}+t_{ab}\,.\label{eq:g4intermsofg0}
\end{eqnarray}
In here the metric $g_{(0)ab}$ is a representative of the conformal class of boundary metrics and $t_{ab}$ is proportional to the boundary stress energy tensor. The trace and divergence of $t_{ab}$ are given by
\begin{equation}
t^a{}_a=-\frac{1}{2}a_{(4)}\,,\qquad\text{and}\qquad\nabla^{(0)a}t_{ab}=0\,,
\end{equation}
where $a_{(4)}$ is the anomaly density 
\begin{equation}
 a_{(4)}=-\frac{1}{8}R_{(0)ab}R^{ab}_{(0)}+\frac{1}{24}R^2_{(0)}\,.
\end{equation}
Indices on $g_{(2)ab}$, $t_{ab}$ etc. are raised with $g_{(0)}^{ab}$. The coefficients of all the higher order terms $\mathcal{O}(z^5)$ and beyond are determined in terms of both $g_{(0)ab}$ and $g_{(4)ab}$. For AAdS$_5$ space-times $g_{(0)ab}$ is conformally flat.

In order to identify the class of FG coordinate systems that are suitable for the study of ASch$_5$ space-times let us first consider the case of pure AdS$_5$. In this case the FG expansion for $h_{ab}$ terminates at the order $z^2$ and its coefficient in this case is given by \cite{Skenderis:1999nb}
\begin{equation}
g_{(4)ab}=\frac{1}{4}g_{(2)ac}g_{(2)b}^c\,.
\end{equation}
Imagine that we are in a Poincar\'e coordinate system for which
\begin{equation}\label{eq:PoincareAdS}
ds^2=\frac{dz^2}{z^2}+\frac{1}{z^2}\left(-2dTdV+(dx^i)^2\right)\,,
\end{equation}
so that $g_{(2)ab}=g_{(4)ab}=0$. In this case we know that performing a TsT transformation with the shift in the V-direction leads to a pure Schr\"odinger space-time in Poincar\'e coordinates. We will use this fact to construct a family of FG coordinate systems that can be used to study the boundary of a pure Sch$_5$ space-time. For this purpose we remind the reader about Penrose--Brown--Henneaux (PBH) coordinate transformations \cite{Penrose:1986ca,Brown:1986nw,Imbimbo:1999bj}. A PBH transformation is a diffeomorphism
\begin{equation}
(z,T,V,x^i)\rightarrow(z',T',V',x'^i)\,,
\end{equation}
where the new coordinates $(z',T',V',x'^i)$ are such that in the primed coordinate system the metric is again of the FG form
\begin{equation}
ds^2=\frac{dz'^2}{z'^2}+\frac{1}{z'^2}\left(g'_{(0)ab}+z'^2g'_{(2)ab}+z'^4g'_{(4)ab}\right)dx'^adx'^b\,,
\end{equation}
where $g'_{(0)ab}=e^{2\sigma}g_{(0)ab}$ and where $g'_{(2)ab}$ and $g'_{(4)ab}$ depend on $\sigma$ and $g_{(0)ab}$ as well as their derivatives. For us the precise form of these transformations is not important (see \cite{Imbimbo:1999bj,Skenderis:2000in} for explicit formulas). The full class of FG coordinate systems can be obtained by starting with the Poincar\'e coordinate system and act on it with an arbitrary PBH transformation. Below we will formulate restrictions that are suitable to the study of Sch$_5$ (and later of ASch$_5$) space-times.

In the TsT transformation we identify the scalar $\Phi$ with some function of the metric component $g_{VV}$, see \eqref{eq:TsTscalarbody}. If we perform an arbitrary coordinate transformation then in general in the new coordinate system $\Phi'(x')=\Phi(x)$ will not depend only on $g'_{V'V'}$. The identification of $\Phi'$ with $g'_{V'V'}$ is preserved under the restricted set of coordinate transformations that satisfy $g_{VV}=g'_{V'V'}$, i.e.
\begin{equation}\label{eq:TsTcoordtrafo}
V = V'+f(z',T',x'^i)\,,\qquad x^A = f^A(z',T',x'^i)\,,
\end{equation}
where $x^A$ is any coordinate other than $V$. These coordinate transformations have the property that $N=\partial_V=\partial_{V'}$. Therefore the TsT formulas \eqref{eq:TsTvectorbody}, \eqref{eq:TsTscalarbody} and \eqref{eq:TsTmetricbody} apply to all coordinate systems that we can get to by starting with Poincar\'e AdS$_5$ and applying the coordinate transformation \eqref{eq:TsTcoordtrafo} to it. This classifies all possible coordinate systems in which we can use adapted coordinates to perform the TsT transformation. If we additionally choose a time slicing and consider diffeomorphisms of the type \eqref{eq:TsTcoordtrafo} that preserve the time slicing then the resulting class of coordinate transformations is contained in the class of double foliation preserving diffeomorphisms of \cite{Horava:2009vy}.

Even though we do not have a coordinate independent way of performing the TsT operations we expect that the result can be written in a coordinate independent manner as follows\footnote{For example the TsT of global AdS (taken to mean any coordinate system that does not depend on global AdS time) can be done as follows. First go to an adapted coordinate system such as Poincar\'e AdS in light cone coordinates, perform the TsT transformation and apply to the TsT transformed metric (Poincar\'e Schr\"odinger) the inverse of the coordinate transformation from global to Poincar\'e AdS. This will give rise to a time-dependent Schr\"odinger coordinate system that can be thought of as the TsT of global AdS. The time dependence is a consequence of the fact that $N$ and the global AdS Hamiltonian do not commute.}
\begin{eqnarray}
 \bar g_{\mu\nu} & = & e^{-2\Phi/3}\left(g_{\mu\nu}-e^{-2\Phi}A_\mu A_\nu\right)\,,\label{eq:TsTmetricgen}\\
 A_\mu & = & \gamma e^{2\Phi}g_{\mu\rho}N^\rho\,,\label{eq:TsTvectorgen}\\
 e^{-2\Phi} & = & 1+\gamma^2 N^\mu N^\nu g_{\mu\nu}\,,\label{eq:TsTscalargen}
\end{eqnarray}
where $N^\mu$ is an AdS$_5$ Killing vector which asymptotically commutes with the 
Schr\"odinger algebra.

We will restrict to FG coordinate systems that have the property that $N^z=0$ so that $N$ is a boundary Killing vector. When $N^z\neq 0$ we cannot make $N^\mu$ manifest in a FG coordinate system and therefore cannot directly perform the TsT transformation. We could circumvent this problem by first performing a PBH transformation to a coordinate system in which $N^{z'}=0$, perform the TsT transformation and subsequently the inverse of the PBH transformation that we did in the beginning. On the other hand it is physically plausible that $N$ should be a boundary Killing vector after TsT (and therefore also before), but without an independent definition of the Schr\"odinger boundary we cannot claim that it must be so. In any case for the applications considered here it will suffice to take $N^z=0$.

The family of FG coordinate systems for pure AdS$_5$ that has $N^z=0$ for an $N$ whose commutant is the Schr\"odinger algebra can be obtained by starting with \eqref{eq:PoincareAdS} and applying to it any PBH transformation with a function $\sigma$ that does not depend on $V$. We will next discuss the properties of these FG coordinate systems. We have
\begin{equation}
 g_{(0)VV} = 0\,,\qquad\partial_V g_{(0)ab}=0\,.\label{eq:bdrymetric1}
\end{equation}
These properties are preserved under conformal rescalings that do not depend on $V$. The family of conformally flat boundary metrics $g_{(0)ab}$ each element of which possesses a null Killing vector, $\partial_V$, can be parametrized by
\begin{equation}\label{eq:bdrymetric}
 g_{(0)ab}dx^adx^b=\text{exp}[2\sigma(T,x^i)]\left(-2dTdV+\delta_{ij}dx^idx^j\right)\,.
\end{equation}

Next we show that all metrics within this conformal class have the property that
\begin{equation}\label{eq:bdrymetric2}
g_{(2)VV} = 0\,.
\end{equation}
Because $g_{(2)VV}$ is conformally invariant under the above restricted class of conformal rescalings that do not depend on $V$ to prove \eqref{eq:bdrymetric2} one simply notes that it vanishes for say the Poincar\'e boundary metric. Further, it also follows that
\begin{equation}\label{eq:bdrymetric3}
g_{(2)Vc}g^c_{(2)V}=0\,.
\end{equation}

If we next consider AAdS$_5$ space-times then what changes is the form of $g_{(4)ab}$ plus the fact that there will appear terms of order $z^3$ and higher in $h_{ab}$. Using the general expression for $g_{(4)ab}$ given in \eqref{eq:g4intermsofg0} as well as \eqref{eq:bdrymetric1}, \eqref{eq:bdrymetric2} and \eqref{eq:bdrymetric3} we see that $g_{(4)VV}$ is given by 
\begin{equation}
g_{(4)VV}=t_{VV}\,,
\end{equation}
so that $g_{(4)VV}$ is fully given in terms of the $VV$ component of the AdS boundary stress tensor. This implies that in the FG coordinates we have
\begin{eqnarray}
 A^\mu & = & \gamma\delta_V^\mu-\gamma^3z^2t_{VV}\delta^\mu_V+\mathcal{O}(z^3)\,,\label{eq:falloffvector}\\
\Phi & = & -\frac{1}{2}\gamma^2z^2t_{VV}+\mathcal{O}(z^3)\,.\label{eq:falloffscalar}
\end{eqnarray}

\section{TsT invariance}

\subsection{The on-shell action}\label{sec:holographicrenorm}

In the previous section we have identified a conformal class of Schr\"odinger boundaries which can be thought of as the TsT image of that subclass of AdS boundaries admitting a null Killing vector. In this section we will study all possible counterterms that are relevant for those ASch$_5$ solutions that can be written as the TsT transformation of an AAdS$_5$ solution. We restrict the analysis to demanding finiteness of the on-shell action. 

In the previous section we denoted the 5-dimensional TsT transformed metric by $\bar g_{\mu\nu}$. Here we will continue to use this notation. Barred expressions such as $\bar R$ are used to indicate that the metric $\bar g$ is involved. However we warn the reader that we do not always carefully distinguish when an index is raised with $\bar g^{\mu\nu}$ or with $g^{\mu\nu}$. It will be clear from the context which inverse metric has been used.

Consider the action \eqref{eq:5Daction} which we repeat here\footnote{We use the following two conventions for the Riemann tensor and
the extrinsic curvature
\begin{eqnarray}
R^\mu{}_{\nu\rho\sigma}
& = & \partial_\rho\Gamma^\mu_{\nu\sigma}+\ldots\,, \\
K_{\mu\nu} & = &
h_\mu{}^\rho\nabla_{\rho}n_{\nu}\,,
\end{eqnarray}
where $h_\mu{}^\rho=\delta_\mu^\rho-n_\mu n^\rho$ and where
$n^\mu$ is the outward pointing unit vector at the
timelike boundary.}
\begin{eqnarray}
 I_{\text{bulk}}+I_{\text{GH}} & = & \frac{1}{16\pi G_N}\int_{M}d^5x\sqrt{-\bar g}\left(\bar R-\frac{4}{3}\partial_\mu\Phi\partial^\mu\Phi-V(\Phi)-\right.\nonumber\\
&&\left.\frac{1}{4}e^{-\tfrac{8}{3}\Phi}F_{\mu\nu}F^{\mu\nu}-4A_\mu A^\mu\right)+\frac{1}{8\pi G_N}\int_{\partial\mathcal{M}}d^4\xi\sqrt{-\bar h}\bar K\,,\nonumber\\
&&\label{eq:5Dactionbody}
\end{eqnarray}
where the potential $V$ is given by
\begin{equation}
 V(\Phi)=4e^{\tfrac{2}{3}\Phi}\left(e^{2\Phi}-4\right)\,.
\end{equation}
In this section we will consider adding counterterms so that the on-shell value of \eqref{eq:5Dactionbody} evaluated on \eqref{eq:TsTvectorbody} to \eqref{eq:TsTmetricbody} is finite. We will show that the form of the counterterms is constrained by the fact that the on-shell value of \eqref{eq:5Dactionbody} (using a cut-off boundary to make the result finite) is TsT invariant as will be proven below. We then use Fefferman--Graham coordinates with $N$ a boundary Killing vector for the AAdS$_5$ metric $g_{\mu\nu}$ and the form of the solutions \eqref{eq:TsTvectorbody} to \eqref{eq:TsTmetricbody} to construct the counterterms.

The action \eqref{eq:5Dactionbody} evaluated for the class of solutions \eqref{eq:TsTvectorbody} to \eqref{eq:TsTmetricbody} is TsT invariant. This means that on-shell we have
\begin{align}
 &\frac{1}{16\pi G_N}\int_{M}d^5x\sqrt{-\bar g}\left(\bar R-\frac{4}{3}\partial_\mu\Phi\partial^\mu\Phi-V(\Phi)-\right.\nonumber\\
&\left.\frac{1}{4}e^{-\tfrac{8}{3}\Phi}F_{\mu\nu}F^{\mu\nu}-4A_\mu A^\mu\right)+\frac{1}{8\pi G_N}\int_{\partial\mathcal{M}}d^4\xi\sqrt{-\bar h}\bar K\nonumber\\
=\nonumber\\
&\frac{1}{16\pi G_N}\int_{\mathcal{M}}d^5 x\sqrt{-g}\left(R+12\right)
+\frac{1}{8\pi G_N}\int_{\partial\mathcal{M}}d^4\xi\sqrt{-h}K\,.\label{eq:TsTinvarianceaction}
\end{align}
This can be shown using the formulas given at the end of appendix \ref{app:TsT}.

The result \eqref{eq:TsTinvarianceaction} has the following important consequence. Whatever the complete set of counterterms for the full space of ASch$_5$ solutions is, upon substituting the solutions \eqref{eq:TsTvectorbody} to \eqref{eq:TsTmetricbody}, the counterterms must cancel the divergences coming from the right hand side of \eqref{eq:TsTinvarianceaction}. This means that counterterms that contribute to the on-shell action divergently must equal, upon use of \eqref{eq:TsTvectorbody} to \eqref{eq:TsTmetricbody}, the counterterms of the usual AAdS$_5$ action \cite{Balasubramanian:1999re}
\begin{eqnarray}
 I_{\text{AAdS$_5$}} & = & \frac{1}{16\pi
G_N}\int_{\mathcal{M}}d^5x\sqrt{-g}\left(R+12\right)\nonumber\\
&&+\frac{1}{8\pi G_N}\int_{\partial\mathcal{M}}d^4\xi\sqrt{-h}
\left(K-3-\frac{1}{4}R_{(h)}\right)\,.\label{eq:AdS5action2body}
\end{eqnarray}

We will write down the most general expression consisting of terms that can contribute to the on-shell action if we evaluate them using \eqref{eq:TsTvectorbody} to \eqref{eq:TsTmetricbody} together with a FG coordinate system for $g_{\mu\nu}$ in which $N$ is a boundary Killing vector. This will suffice for the purposes of this paper, but for completeness we will indicate at the end of this subsection what kind of terms could be added when $N$ is not a boundary Killing vector. We denote by $I$ the action
\begin{equation}
 I=I_{\text{bulk}}+I_{\text{GH}}+I^{(0)}_{\text{ct}}+I^{(2)}_{\text{ct}}+I_{\text{ext}}\,,
\end{equation}
where $I_{\text{bulk}}+I_{\text{GH}}$ is given in \eqref{eq:5Dactionbody}, $I^{(0)}_{\text{ct}}+I^{(2)}_{\text{ct}}$ contains only intrinsic counterterms without derivatives ($I_{\text{ct}}^{(0)}$) or terms that are second order in derivatives ($I_{\text{ct}}^{(2)}$) and where $I_{\text{ext}}$ consists of extrinsic counterterms for the massive vector field $A_\mu$ and scalar $\Phi$. We find for $I^{(0)}_{\text{ct}}$, $I^{(2)}_{\text{ct}}$ and $I_{\text{ext}}$
\begin{eqnarray}
 I^{(0)}_{\text{ct}} & = & \frac{1}{8\pi G_N}\int_{\partial\mathcal{M}}d^4\xi\sqrt{-\bar h}
 \left(a_0+a_1\Phi+a_2\Phi^2+a_3A_aA^a+a_4\Phi A_aA^a\right.\nonumber\\
&&\left.+a_5(A_aA^a)^2\right)\,,\\
I^{(2)}_{\text{ct}} & =  & \frac{1}{8\pi G_N}\int_{\partial\mathcal{M}}d^4\xi\sqrt{-\bar h}\left(\left(b_0+b_1\Phi+b_2A_aA^a\right)\bar R_{(\bar h)}\right.\nonumber\\
&&\left.+ b_3F_{ab}F^{ab}\right)\,,\\
I_{\text{ext}} & = & \frac{1}{8\pi G_N}\int_{\partial\mathcal{M}}d^4\xi\sqrt{-\bar h}\left(\left(c_0+c_1\Phi+c_2A_bA^b\right)\bar n^\mu A^aF_{\mu a}\right.\nonumber\\
&&\left.+\left(c_3+c_4\Phi+c_5A_aA^a\right)\bar n^\mu\partial_\mu\Phi\right)\,.
\end{eqnarray}
The counterterms in $I^{(0)}_{\text{ct}}$ have also been considered in \cite{Herzog:2008wg,Adams:2008wt,Ross:2009ar}. Some of the extrinsic counterterms have also been considered in \cite{Adams:2008wt}.

Since $\sqrt{-\bar h}$ is $\mathcal{O}(z^{-4})$ we only need to include counterterms that are at most of order $z^4$. In order to see that $I^{(0)}_{ct}$ contains all possible terms note that for our class of solutions both $\Phi$ and $A_aA^a$ are $\mathcal{O}(z^2)$ near $z=0$. To show that $I^{(2)}_\text{ct}$ consists of all possible terms that are nonvanishing for our class of solutions we argue as follows. In general $R_{(h)}$ is an order $z^2$ term. The extra conditions \eqref{eq:bdrymetric1} do not change that. This means that also $\bar R_{(\bar h)}$ evaluated for the TsT transformed solution, as given in \eqref{eq:TsTRh}, will be $\mathcal{O}(z^2)$. The term $F^2$ evaluated for an arbitrary TsT transformed solution is of order $z^0$, but with the use of \eqref{eq:bdrymetric1}, i.e. for the case of a TsT transformation with a null Killing vector that is also a boundary null Killing vector, it can be seen to be of order $z^4$. A term such as $\left(\bar\nabla^{(\bar h)}_aA_b+\bar\nabla^{(\bar h)}_bA_a\right)^2$ is $\mathcal{O}(z^6)$ and so it does not contribute to the on-shell action. Terms such as $\bar R_{(\bar h)ab}A^aA^b$ are not included because they are by partial integration equivalent to a combination of $F^2$ and $\left(\bar\nabla^{(\bar h)}_aA_b+\bar\nabla^{(\bar h)}_bA_a\right)^2$ plus a term of the form $(\bar\nabla^{(\bar h)}_aA^a)^2$ which does not contribute. Further a term such as $\bar h^{ab}\partial_a\Phi\partial_b\Phi$ is $\mathcal{O}(z^6)$ and so will never contribute to the on-shell action.

The term $I_{\text{ext}}$ has been added for the following reason. The TsT identity \eqref{eq:TsTinvarianceaction} suggests that the only relevant extrinsic counterterm for our class of solutions is the GH boundary term. However, due to TsT relations such as \eqref{eq:TsTidentity1} and \eqref{eq:bulkTsT6} it is possible to write down extrinsic counterterms for both the vector field and the scalar field which are such that they cancel divergent terms among themselves (we will confirm later that this must be so) rather then with the bulk action. When this is the case they can still contribute to the on-shell action by a finite amount and secondly they may be very relevant for considerations related to the variation of the action since the extrinsic counterterms for the massive vector and the scalar will not cancel each other after variation.

Evaluating $I_{\text{ct}}^{(0)}$, $I_{\text{ct}}^{(2)}$ and $I_{\text{ext}}$ for the TsT transformed solutions we obtain
\begin{eqnarray}
I_{\text{ct}}^{(0)} & = & \frac{1}{8\pi G_N}\int_{\partial\mathcal{M}}d^4\xi\sqrt{-h}\left(a_0+\left(-\frac{a_0}{3}+a_1-2a_3\right)\Phi\right.\nonumber\\
&&\left.+\frac{1}{18}\left(a_0-6a_1+18a_2-48a_3-36a_4+72a_5\right)\Phi^2\right)\,,\\
I^{(2)}_{\text{ct}} & = & \frac{1}{8\pi G_N}\int_{\partial\mathcal{M}}d^4\xi\sqrt{-h}\left[b_0R_{(h)}+\left(\frac{b_0}{3}+b_1-2b_2\right)\Phi R_{(h)}\right.\nonumber\\
&&\left.+\left(b_3+\frac{1}{4}b_0\right)F_{ab}F^{ab}\right]\,,\\
I_{\text{ext}} & = & \frac{1}{8\pi G_N}\int_{\partial\mathcal{M}}d^4\xi\sqrt{-h}\left((c_3-2c_0)n^\mu\partial_\mu\Phi\right.\nonumber\\
&&\left.+2\left(2c_1-4c_2+\frac{16}{3}c_0-c_4+2c_5\right)\Phi^2\right)\,,
\end{eqnarray}
where one should think of the integrands as being expanded up to $\mathcal{O}(z^0)$. This will involve the FG expansion coefficients $g_{(5)VV}$ and $g_{(6)VV}$ appearing in the expansion of $\Phi$ at the orders $z^3$ and $z^4$, respectively. 

In order to have equality with the divergent terms in the boundary action of \eqref{eq:AdS5action2body} we need
\begin{eqnarray}
a_0 & = & -3\,,\\
b_0 & = & -\frac{1}{4}\,,\\
c_3-2c_0 & = & 0\,,\label{eq:c1}\\
-\frac{a_0}{3}+a_1-2a_3 & = & 0\,.\label{eq:a1}
\end{eqnarray}
The $\gamma$ independent divergent terms must be equal to the corresponding counterterms of \eqref{eq:AdS5action2body}. This fixes the $a_0$ and $b_0$ coefficients. The $\gamma$ dependent divergent terms must cancel among themselves. There are two such terms $\Phi$ and $n^\mu\partial_\mu\Phi$. Assuming that in general $g_{(5)VV}\neq 0$ these cannot cancel each other so their coefficients must vanish independently. This leaves us with finite terms proportional to $\gamma$. Demanding that the counterterms reduce to the counterterms of \eqref{eq:AdS5action2body} we set these terms to zero as well. This gives\footnote{Later on when we discuss black hole solutions we will introduce the grand potential. This quantity involves the difference of the (Euclidean) on-shell action for a black hole space-time and some background space-time such as pure Schr\"odinger. This could therefore be computed using a background subtraction method. In order that the background subtraction method gives the same answer as the difference of two renormalized on-shell actions we need to impose \eqref{eq:TsTinvonshellaction} to \eqref{eq:b3}.}
\begin{eqnarray}
0 & = & \frac{1}{18}\left(a_0-6a_1+18a_2-48a_3-36a_4+72a_5\right)\nonumber\\
&& +2\left(2c_1-4c_2+\frac{16}{3}c_0-c_4+2c_5\right) \,,\label{eq:TsTinvonshellaction}\\
b_1-2b_2 & = & \frac{1}{12}\,,\label{eq:brelation2}\\
b_3 & = & \frac{1}{16}\,.\label{eq:b3}
\end{eqnarray}

Even though it remains to be seen if by some appropriate definition of ASch$_5$ space-times we should consider the case where $N$ is not a boundary Killing vector we briefly discuss the effect of our assumption that in the FG expansion for $g_{\mu\nu}$ we take $N$ to be a boundary null Killing vector. When we drop this assumption we have $N^z\neq 0$ in which case both $A^2$ and $F^2$ are of order $z^0$ ($\Phi$ is of course still of order $z^2$). This means that we can now take arbitrary powers of $A^2$ and $F^2$ and add those as counterterms\footnote{The formulas \eqref{eq:bdryTsT1} to \eqref{eq:TsTRh} change when $N$ is no longer a boundary Killing vector.}. 

In the AAdS$_5$ case the counterterms of \eqref{eq:AdS5action2body} respect the symmetries of the boundary theory. From a bulk space-time perspective this means that they are invariant under both boundary diffeomorphisms as well as PBH transformations (up to anomalous transformations \cite{Henningson:1998gx}). When performing the TsT transformation with a null Killing vector $N$ we can in principle distinguish two cases: 1). $N$ is a boundary vector and 2). $N$ is not a boundary vector. The first case is naturally selected by the TsT transformation and in this case the local symmetries are boundary diffeomorphisms and (possibly anomalous) PBH transformations that do not depend on $V$. Our point of view has been to restrict to case one without claiming that this is the only possibility and to construct for this case the local counterterms. In \cite{Guica:2010sw} it is suggested that one should allow for counterterms that are nonlocal in the $V$ direction. It would be interesting to explore these directions further.

It should be clear that demanding that there is a well-posed variational problem for $I$ lies outside the scope of this work. Rather, we will focus on thermodynamics for which the above counterterm analysis suffices.

\subsection{Horizon properties}

So far we have considered a rather general class of ASch$_5$ space-times. In the remainder we will focus on asymptotically Schr\"odinger black holes that can be obtained as the TsT transform of a pure AAdS$_5$ black hole and study their thermodynamic properties.

The type of AAdS$_5$ black holes that we will consider have the following properties. The black hole space-time possesses three commuting Killing vectors one of which is $N$ and it possesses a Killing horizon that is generated by the Killing vector $X$ that is a linear combination of these three Killing vectors. The Killing vector $X$ satisfies the usual equations $X^\mu\nabla_\mu X^\nu=\kappa X^\nu$ and $X^\mu X_\mu=0$ on the horizon, where $\kappa$ is the surface gravity. It follows that on the horizon $X$ is also hypersurface orthogonal. Below it will be assumed that $\kappa\neq 0$.

The horizon properties of the TsT transformed black hole are inherited from the metric $g_{\mu\nu}$. To see this note that, as follows from \eqref{eq:KillingTsT}, after TsT all three Killing vectors, and thus in particular $X$, remain to be Killing vectors. The norm of $X$ with respect to $\bar g_{\mu\nu}$ is given by
\begin{equation}
\vert\vert X\vert\vert^2_{\text{TsT}}=e^{-2\Phi/3}\vert\vert X\vert\vert^2-e^{-8\Phi/3}\left(A_\mu X^\mu\right)^2\,.
\end{equation}
The horizon of the metric $g_{\mu\nu}$ is the hypersurface at which $\vert\vert X\vert\vert^2$ vanishes. Since also $A_\mu X^\mu\propto N_\mu X^\mu$ vanishes there we find that $\vert\vert X\vert\vert^2_{\text{TsT}}$ vanishes at the same locus as does $\vert\vert X\vert\vert^2$. To prove that $N_\mu X^\mu$ vanishes on the horizon one can use hypersurface orthogonality of $X$, the fact that $N$ and $X$ commute and that $X$ is null on the horizon. We conclude that the location of the horizon is preserved by TsT. Finally from the fact that on the horizon
\begin{equation}
X^\mu\bar\nabla_\mu X^\nu=X^\mu\nabla_\mu X^\nu
\end{equation}
it follows that after TsT the horizon is still a Killing horizon that is generated by $X$ and that the surface gravity is TsT invariant.

From equation \eqref{eq:detinducedTsT} it follows that the determinant of the induced metric on the spatial part of the horizon (which is a co-dimension 2 surface having $V$ as one of its tangential directions) is preserved. Hence the horizon area is preserved under TsT (see also \cite{Gimon:2003xk}).

We assumed that the pure AAdS$_5$ space-time before TsT possesses three Killing vectors. Let us denote these as $\partial_T$, $N=\partial_V$ and $\partial_\psi$, where $T$ is some time coordinate. Assume that we Wick rotate the metric before TsT and impose periodicities on $iT$, $iV$ and $i\psi$ such that the Wick rotated metric does not contain any conical singularities. These conical singularities can arise on a 2-dimensional surface that can be obtained as follows. Introduce coordinates that are diagonal in a radial coordinate, $r$ say, such that the horizon is at $g^{rr}=0$. Consider the equation $X^\mu\partial_\mu f=0$ where $f$ is a function of the remaining four coordinates. This equation defines three independent hyperplanes. Consider the induced metric on the common intersection of these three hyperplanes and expand to leading order in $r$. This defines the induced metric on a 2-dimensional surface which upon Wick rotation (and removal of the conical singularity) is known as the bolt \cite{Gibbons:1979xm}, i.e. the fixed point set of the vector $X$. Now let us consider repeating this calculation for the TsT transformed metric $\bar g_{\mu\nu}$. It can be shown that $N_\mu dx^\mu$ vanishes on the bolt (as follows essentially from the property that $N^\mu X_\mu=0$ on the horizon) and hence the induced metric on the bolt conformally rescales under TsT (the rescaling is regular on the horizon as long as the scalar field is regular on the horizon) and therefore the periodicities of $iT$, $iV$ and $i\psi$ needed to remove the conical singularity are TsT invariant. This means that the temperature and chemical potentials associated with a particular choice of Hamiltonian etc. are TsT invariant. This can be thought of as a generalization of the results of \cite{Gimon:2003xk,Horowitz:1993wt}.

In order to do thermodynamics we will use the thermodynamical action formalism of \cite{Brown:1990fk}. The thermodynamical action is the action evaluated for the complexified Wick rotated geometry obtained by setting $T=-i\tau$ in which the integration is from the horizon to the boundary. We compute the thermodynamical action by integrating $I_\text{bulk}+I_{\text{GH}}$ from the horizon to some cut-off boundary and use a minimal subtraction method to regularize it. The difference between the thermodynamical action for a TsT transformed black hole space-time and a TsT transformed space-time at the same temperature and chemical potentials but without the black hole, assuming that our system is well described by a grand canonical ensemble, is the grand potential. From the results of the previous subsection it follows that the grand potential is TsT invariant.

\section{Phase transitions in global Schr\"odinger space-time}

One of the original questions that motivated this research was to find out if the global Schr\"odinger space-time of \cite{Blau:2009gd} in analogy with global AdS would show a Hawking--Page type phase transition \cite{Hawking:1982dh}. An important difference with global AdS is that the spatial sections of the global Schr\"odinger boundary are non-compact. On the other hand in this case we have a harmonic trapping potential so it could still be that there is some critical temperature at which a phase transition occurs from pure thermal Schr\"odinger to an asymptotically Schr\"odinger black hole. Now that we know that the thermodynamics is TsT invariant we can answer this question by looking at the same problem before TsT, i.e. does there exist a phase transition between thermal plane wave AdS and some AAdS black hole space-time? We will show in this section that this is indeed the case. 

In \cite{Maldacena:2008wh} a black hole is constructed that is time independent with respect to a Killing vector that becomes asymptotically the AdS$_5$ plane wave Hamiltonian. This solution is obtained by taking an appropriate scaling limit of a 5-dimensional Kerr black hole in which the angular velocity is sent to the speed of light. The solution is asymptotic to thermal plane wave AdS$_5$ at finite chemical potential. We will discuss the geometric and thermodynamic properties of this black hole and show that there is a phase transition between this black hole and thermal plane wave AdS$_5$ at finite chemical potential. Then upon performing a TsT transformation to both the black hole solution and pure plane wave AdS and using the TsT invariance of thermodynamics we find the sought for phase transition of the global Schr\"odinger space-time.

One may object that we have actually only proven that local thermodynamic quantities are TsT invariant, but that this need not apply to the full saddle point approximation of the partition function. However, the conditions that we impose on the saddle points are such that they always have a Killing vector that becomes asymptotically null and so the saddle points exist both before and after TsT.

In the next subsection we will review and slightly generalize the scaling argument of \cite{Maldacena:2008wh}. After this we will study some of its geometric and thermodynamic properties. At the end of this section we will discuss the phase transition.

\subsection{AAdS$_5$ black holes with an asymptotic null Killing vector}\label{sec:scalings}

Consider the 5-dimensional Kerr-AdS metric \cite{Hawking:1998kw} in coordinates ($\tau,r,\theta,\phi,\psi$) that are stationary at infinity
\begin{eqnarray}
ds^2 & = & -\frac{\Delta_{\theta}^{(a,b)}(1+r^2)}{\Xi_a\Xi_b}d\tau^2 + \frac{2m}{\Sigma^2_{(a,b)}}\left(\frac{\Delta_{\theta}^{(a,b)}d\tau}{\Xi_a\Xi_b}-\omega_{(a,b)} \right)^2+\frac{\Sigma^2_{(a,b)}}{\Delta_r^{(a,b)}}dr^2 \nonumber\\
&& +\frac{\Sigma^2_{(a,b)}}{\Delta_\theta^{(a,b)}}d\theta^2 + \frac{r^2+a^2}{\Xi_a}\cos^2\theta d\phi^2+ \frac{r^2+b^2}{\Xi_b}\sin^2\theta d\psi^2\,,\label{eq:nonrotatingboundary}
\end{eqnarray}
where
\begin{eqnarray}
\omega_{(a,b)} & = & \frac{a\cos^2\theta d\phi}{\Xi_a} + \frac{b\sin^2\theta d\psi}{\Xi_b}\,,\\
\Sigma^2_{(a,b)} & = & r^2+a^2\sin^2\theta+b^2\cos^2\theta\,,\\
\Delta_r^{(a,b)} & = & \frac{1}{r^2}(r^2+a^2)(r^2+b^2)(r^2+1)-2m\,,\\
\Delta_\theta^{(a,b)} & = & 1-a^2\sin^2\theta-b^2\cos^2\theta\,,\\
\Xi_a & = & 1-a^2\,,\\
\Xi_b & = & 1-b^2\,.
\end{eqnarray}
In the expression given in \cite{Hawking:1998kw} we put the AdS radius $l=1$ and replaced $\theta$ by $\tfrac{\pi}{2}-\theta$ following \cite{Maldacena:2008wh}. The ranges of the angular coordinates are $0\le\theta\le\pi/2$, $-\pi\le\phi<\pi$ and $-\pi\le\psi<\pi$. Contrary to the case of asymptotically flat Kerr black holes, the parameters $a$ and $b$ are restricted to $a^2<1$ and $b^2<1$.

Now define the coordinates $T$ and $V$ via
\begin{eqnarray}
\phi & = & T-2(1-a)V\,,\\
\tau & = & T\,,
\end{eqnarray}
and substitute this into \eqref{eq:nonrotatingboundary}. The result is
\begin{eqnarray}
 ds^2 & = & -\left(1+\frac{r^2+b^2}{\Xi_b}\sin^2\theta\right)dT^2-\frac{4}{1+a}(r^2+a^2)\cos^2\theta dTdV\nonumber\\
&&+\frac{2m}{\Sigma^2_{(a,b)}}\left(\frac{1+a\sin^2\theta-b^2\cos^2\theta}{(1+a)\Xi_b}dT+\frac{2a\cos^2\theta}{1+a}dV-\frac{b\sin^2\theta}{\Xi_b}d\psi\right)^2\nonumber\\
&&+\frac{4(1-a)}{1+a}(r^2+a^2)\cos^2\theta dV^2+\frac{\Sigma^2_{(a,b)}}{\Delta_r^{(a,b)}}dr^2+\frac{\Sigma^2_{(a,b)}}{\Delta_\theta^{(a,b)}}d\theta^2\nonumber\\
&&+ \frac{r^2+b^2}{\Xi_b}\sin^2\theta d\psi^2\,.\label{eq:beforescaling}
\end{eqnarray}
Since $\phi$ is periodic with period $2\pi$ we have that $V$ is periodic with period $\tfrac{\pi}{1-a}$. Setting $a=1$ we obtain
\begin{eqnarray}
ds^2 & = & -\left(1+\frac{r^2+b^2}{\Xi_b}\sin^2\theta\right)dT^2-2(r^2+1)\cos^2\theta dTdV\nonumber\\
&&+\frac{m}{2\Sigma^2_{(1,b)}}\left(\frac{1+\sin^2\theta-b^2\cos^2\theta}{\Xi_b}dT+2\cos^2\theta dV-\frac{2b}{\Xi_b}\sin^2\theta d\psi\right)^2\nonumber\\
&&+\frac{\Sigma^2_{(1,b)}}{\Delta_r^{(1,b)}}dr^2+\frac{\Sigma^2_{(1,b)}}{\Xi_b\cos^2\theta}d\theta^2+\frac{r^2+b^2}{\Xi_b}\sin^2\theta d\psi^2\,.\label{eq:scalinglimit}
\end{eqnarray}
In this metric $V$ runs from $-\infty$ to $+\infty$. We can however compactify $V$ by identifying $V\sim V+2\pi L$. After this identification the resulting metric is no longer obtainable from \eqref{eq:nonrotatingboundary} via some scaling limit. Later we will see that, by choosing different coordinates, \eqref{eq:scalinglimit} can be thought of as asymptotically plane wave AdS$_5$ with a compact null coordinate. We will refer to the solution \eqref{eq:scalinglimit} with $b^2<1$ and $a=1$ as the Maldacena--Martelli--Tachikawa (MMT) black hole \cite{Maldacena:2008wh}. The limit $a=1$ can be thought of as a limit in which one rotation approaches the speed of light. We also refer to \cite{Hawking:1998kw} for a discussion of the limit in which the rotation of an AAdS black hole is sent to the speed of light.

We can define a second scaling limit in which we also send the parameter $b$ to one. This can be done by defining the following limit\footnote{An alternative to this scaling limit is to perform the coordinate transformation $\theta=(1-b^2)^{1/2}\rho$ and to set $b=1$ afterwards.}
\begin{eqnarray}
b & = & 1-\frac{1}{2\lambda^2}\,,\\
\theta & = & \frac{\rho}{\lambda}\,,
\end{eqnarray}
and sending $\lambda\rightarrow\infty$. The result is (after defining $R^{-2}=1+r^2$) \cite{Maldacena:2008wh}
\begin{eqnarray}
ds^2 & = & \frac{1}{R^2}\left(-(R^2+\rho^2)dT^2-2dTdV+d\rho^2+\rho^2d\psi^2\right)\label{eq:secondscaling}\\
&&+\frac{1}{1-2m(R^4-R^6)}\frac{dR^2}{R^2}+\frac{m}{2}R^2\left(dT+2dV-2\rho^2(d\psi-dT)\right)^2\,.\nonumber
\end{eqnarray}

Both the metric \eqref{eq:scalinglimit} and \eqref{eq:secondscaling} are asymptotically plane wave AdS$_5$ and will after TsT be asymptotically global Schr\"odinger.

The difference between the scaling limits to $a=1$ and to $b=1$ is that in the former case $V$ can be compactified while in the latter case $\rho$ cannot. This shows itself in the fact that for \eqref{eq:secondscaling} the mass density is finite but the total mass is infinite while for \eqref{eq:scalinglimit} the total mass is finite.

Finally we can apply even a third scaling limit by starting with \eqref{eq:secondscaling} and defining
\begin{equation}
R =\frac{\bar r}{\lambda}\,,\quad T = \frac{t}{\lambda}\,,\quad V = \frac{\xi}{\lambda}\,,\quad \rho = \frac{\rho}{\lambda}\,,\quad m = \lambda^4\bar m\,.
\end{equation}
Then after sending $\lambda\rightarrow\infty$ we obtain the AdS black hole (brane) solution that is asymptotically Poincar\'e AdS and whose metric reads
\begin{equation}
 ds^2=\frac{1}{\bar r^2}\left(-2dtd\xi+d\rho^2+\rho^2d\psi^2\right)+\frac{\bar m}{2}\bar r^2(dt+2d\xi)^2+\frac{d\bar r^2}{\bar r^2\left(1-2\bar m\bar r^4\right)}\,.\label{eq:blackbrane1}
\end{equation}
This solution is asymptotically Poincar\'e AdS$_5$ and after TsT (shifting along $\xi$) it gives rise to an asymptotically Poincar\'e Schr\"odinger black brane. The TsT-transform of this black brane solution has been studied in \cite{Maldacena:2008wh,Herzog:2008wg,Adams:2008wt,Imeroni:2009cs}.

\subsection{Geometric properties of the MMT black hole}\label{subsec:geometricproperties}

In this subsection we will collect some geometric properties of the MMT black hole \eqref{eq:scalinglimit} that are relevant for its thermodynamic properties.

The horizon at $r=r_H$ is given by the largest positive root of the equation $g^{rr}=0$, i.e.
\begin{equation}\label{eq:horizonb}
(r^2+b^2)(r^2+1)^2-2mr^2 = 0\,.
\end{equation}
A necessary and sufficient condition for the existence of a horizon is provided by
\begin{equation}\label{eq:condition2}
 m\ge \frac{1}{16}\left(20b^2+8-b^4+\vert b\vert(b^2+8)^{3/2}\right)\,.
\end{equation}
When \eqref{eq:condition2} is a strict inequality there will be an inner and an outer horizon. In the extremal case, i.e. when \eqref{eq:condition2} is an equality, the outer horizon is infinitely far away from any point in the space-time. It then looks as if both horizons have coalesced at
\begin{equation}
 r_H^2=-\frac{b^2}{4}+\frac{\vert b\vert}{4}(b^2+8)^{1/2}\,.
\end{equation}
This is the lowest possible value $r_H$ can attain. When $b=0$ there is only one positive real root given by $r=r_H\equiv\sqrt{-1+\sqrt{2m}}$ assuming that $2m>1$.

Besides an event horizon at $g^{rr}=0$ the metric also has a stationary limit surface at $g_{TT}=0$. When $g_{TT}>0$ a massive particle can no longer stand still, i.e. the worldline whose tangent is proportional to $\tfrac{\partial}{\partial T}$ is no longer timelike.

To see that the MMT black hole is asymptotically plane wave AdS$_5$ perform the following coordinate transformation
\begin{eqnarray}
1+r^2 & = & \frac{1}{2R^2}\left(1+\Xi_b(\rho^2+R^2)+F(R,\rho)\right)\,,\label{eq:Rrhowithb1} \\
\cos^2\theta & = & \frac{1}{2\Xi_b R^2}\left(1+\Xi_b(\rho^2+R^2)-F(R,\rho)\right)\,,\label{eq:Rrhowithb2}
\end{eqnarray}
where
\begin{equation}
F(R,\rho) = \sqrt{(1+\Xi_b(\rho^2-R^2))^2+(2\Xi_b\rho R)^2}\,,
\end{equation}
whose inverse is
\begin{eqnarray}
 \frac{1}{R^2} & = & (1+r^2)\cos^2\theta\,,\\
\frac{\rho^2}{R^2} & = & \frac{r^2+b^2}{\Xi_b}\sin^2\theta\,.
\end{eqnarray}
In this coordinate system the metric \eqref{eq:scalinglimit} is asymptotically 
\begin{eqnarray}
ds^2 &=& \frac{1}{R^2}\left(-(\rho^2+R^2)dT^2-2dTdV + d\rho^2 + \rho^2d\psi^2+dR^2 \right)+ \frac{2mR^2dR^2}{(1+\Xi_b\rho^2)^2} \nonumber\\
&& + \frac{mR^2}{2(1+\Xi_b\rho^2)^3} \left((1+2\rho^2)dT+2dV-\frac{}{}2b\rho^2d\psi\right)^2  + \mathcal{O}(R^3)\,.\label{eq:expansionrotatingMMT}
\end{eqnarray}

In order to study the metric near the horizon we define the Eddington--Finkelstein (EF) coordinates $u_{\mp}$, $v_\mp$ and $w_\mp$ via
\begin{eqnarray}
 du_\mp & = & dT\pm\frac{(r^2+b^2)(r^2+1)}{r^2\Delta_r^{(1,b)}}dr\,,\label{eq:EFu}\\
 dv_\mp & = & dV\pm\frac{(r^2+b^2)(r^2-1)}{2r^2\Delta_r^{(1,b)}}dr\,,\label{eq:EFv} \\
 dw_\mp & = & d\psi\pm\frac{b(r^2+1)^2}{r^2\Delta_r^{(1,b)}}dr\,,\label{eq:EFw}
\end{eqnarray}
where each constant set of $(u_+,v_+,w_+)$ values describes an
outgoing null geodesic while each constant set of $(u_-,v_-,w_-)$ values describes an ingoing null geodesic. In these EF coordinates the metric \eqref{eq:scalinglimit} takes the form
\begin{eqnarray}
 ds^2 & = & g_{TT}du_\mp^2+2g_{TV}du_\mp
dv_\mp+2g_{T\psi}du_\mp dw_\mp +g_{VV}dv_\mp^2
 \nonumber\\ &&+2g_{V\psi}dv_\mp dw_\mp +g_{\psi\psi}dw_\mp^2 \pm
2\cos^2\theta drdv_\mp  \nonumber\\
&&\pm\left(1+\frac{1+b^2}{\Xi_b}\sin^2\theta\right)drdu_\mp \mp \frac{2b}{\Xi_b}\sin^2\theta drdw_\mp+g_{\theta\theta}d\theta^2\,.
\end{eqnarray}

Slices of constant $r$, including the horizon at $r=r_H$, contain a non-compact direction parametrized by $\theta$. This is not so in the case of the 5-dimensional Kerr-AdS black hole \eqref{eq:nonrotatingboundary} with $a^2<1$ whose spatial sections have a compact horizon topology that is homeomorphic to a 3-sphere. When we set $a=1$ in \eqref{eq:beforescaling} two things change: i). The $m$-independent term proportional to $dV^2$ in \eqref{eq:beforescaling} vanishes, so that for large $r$ the coordinate $V$ becomes a null coordinate. Secondly, because of the behavior of $\Delta^{(a,b)}_\theta$ as $a$ goes to one the physical distance from a point $P=(T_0,V_0,r_0,\theta_0,\psi_0)$ to a point $Q=(T_0,V_0,r_0,\pi/2,\psi_0)$ diverges. Therefore we can find two points on the horizon $r=r_H$ whose physical distance is infinite. This is not possible for $a^2<1$. After setting $a=1$ and compactifying the $V$ coordinate we can think of the spatial sections of the horizon as a surface that is homeomorphic to a 3-sphere that has been infinitely stretched in the $\theta$ direction. This can be confirmed by studying the metric in EF coordinates on slices of constant $u_-$ and $r=r_H$. Despite the fact that the horizon contains a non-compact direction the area of the horizon is finite and given by
\begin{equation}\label{eq:horizonarea}
A=\frac{2\pi^2L}{\Xi_b}\frac{(r_H^2+1)(r_H^2+b^2)}{r_H}\,.
\end{equation}

The normal to the horizon, the gradient $g^{\mu\nu}\partial_\nu r$, is
proportional to the Killing vector $X$ given by
\begin{equation}
 X=\frac{\partial}{\partial T}-\mu_N L\frac{\partial}{\partial V}-\Omega\frac{\partial}{\partial\psi}\,.
\end{equation}
When $r_H\ge 1$ the norm of $X$ is timelike outside the horizon. When $0<r_H<1$ we numerically checked that there exists a large region strictly outside the horizon (not overlapping with the horizon) that extends all the way to the boundary where the norm of $X$ is spacelike.

Working in the EF coordinate system we can compute the surface gravity $\kappa$ via
\begin{equation}\label{eq:surfacegravity}
 X^\mu\nabla_\mu X^\rho=\kappa X^\rho\,,
\end{equation}
evaluated on the horizon. The result is that
\begin{equation}
 \kappa=\frac{1}{r_H}\frac{2r_H^4+b^2(r_H^2-1)}{r_H^2+b^2}\,.
\end{equation}

In order to avoid conical singularities upon Wick rotating $T=-i\tau$ we must make the identifications
\begin{eqnarray}
\tau & \sim & \tau+\beta\,,\\
iV &\sim& iV + \beta L\mu_N\,, \label{eq:identificationImV}\\
i\psi &\sim& i\psi + \beta\Omega \,,\label{eq:identificationImpsi}
\end{eqnarray}
where the inverse temperature $\beta$ and chemical potentials $\mu_N$ and $\Omega$ are
\begin{eqnarray}
 \beta&=&\frac{2\pi r_H(r_H^2+b^2)}{2r_H^4+b^2(r_H^2-1)}\,,\label{eq:invtemp}\\
\mu_N&=& -\frac{1}{2L}\frac{r_H^2-1}{(r_H^2+1)}\,,\\
\Omega&=&-\frac{b(r_H^2+1)}{r_H^2+b^2}\,.
\end{eqnarray}
The temperature has no local extrema in particular $\partial\beta/\partial r_H<0$.

\subsection{Thermodynamics}\label{app:charges}

The Euclidean action or thermodynamical action defined by
$iI=-I_E$ after Wick rotating $T=-i\tau$ is given by
\begin{equation}\label{eq:Euclideanaction}
 I_E=-\frac{1}{16\pi
G_N}\int_{\mathcal{M}}d^5x\sqrt{g}\left(R+12\right)-
\frac{1}{8\pi
G_N}\int_{\partial\mathcal{M}}d^4\xi\sqrt{h}\left(K-3
-\frac{1}{4}R_{(h)}\right)\,.
\end{equation}
The Wick rotation $T=-i\tau$ leads to a complex geometry for which the above Euclidean action is real-valued. The action is integrated from the horizon to the boundary. The Wick rotated (complex) geometry is a saddle point of this action.

We define $\Delta I_E$ as the difference between the renormalized on-shell action evaluated for the MMT black hole in a given coordinate system and the renormalized on-shell action evaluated for the pure AdS$_5$ metric obtained by setting $m=0$ in that coordinate system. Even though the on-shell action transforms anomalously under PBH transformations \cite{Henningson:1998gx} the difference $\Delta I_E$ is the same in all coordinate systems \cite{Papadimitriou:2005ii}. Assuming that we can treat the thermodynamics as the infinite volume limit of a grand canonical ensemble the grand potential $\Phi_G$ is given by $\Phi_G=\beta^{-1}\Delta I_E$. We find for $\Delta I_E$ the following expression
\begin{equation}
\Delta I_E = -\frac{\pi L\beta}{8\Xi_b G_N}\frac{(r_H^4-1)(r_H^2+b^2)}{r_H^2}\,.
\end{equation}

The difference $\Delta I_E$ satisfies the usual quantum statistical relations
\begin{eqnarray}
\left(\frac{\partial\Delta I_E}{\partial\beta}\right)_{\mu_N,\Omega} & = & E-\mu_N N-\Omega J\,,\label{eq:qstat1}\\
\left(\frac{\partial\Delta I_E}{\partial\Omega}\right)_{\beta,\mu_N} & = & -\beta J\,,\\
\left(\frac{\partial\Delta I_E}{\partial\mu_N}\right)_{\beta,\Omega} & = & -\beta N\,,\label{eq:qstat3}
\end{eqnarray}
where $E$ is the energy, $N$ the momentum conjugate to $\mu_N$ and $J$ the angular momentum conjugate to $\Omega$. The on-shell action $\Delta I_E(\beta,\mu_N,\Omega)$ has no stationary points.

The quantities $E$, $N$ and $J$ can also be expressed in terms of the Brown--York (BY) charges $Q_{\partial_T}$, $Q_{\partial_V}$ and $Q_{\partial_\psi}$ as
\begin{eqnarray}
E & = & Q_{\partial_T}-Q_{\partial_T}(\text{$m=0$ solution})\,,\\
N & = & LQ_{\partial_V}\,,\\
J & = & Q_{\partial_\psi}\,,
\end{eqnarray}
where $Q_{\partial_T}(\text{$m=0$ solution})$ is the Casimir energy computed in a coordinate system that has the same conformal boundary as the coordinate system in which $Q_{\partial_T}$ is computed. The conserved BY charges $Q_\chi$ associated with a Killing vector $\chi$ that is also a boundary Killing vector are defined as \cite{Brown:1992br,Balasubramanian:1999re}
\begin{equation}\label{eq:BYcharge}
 Q_\chi=\int_{\Sigma}d\Sigma\sqrt{\sigma}
u^a\chi^bT_{ab}\,,
\end{equation}
where the BY stress tensor is given by
\begin{equation}\label{eq:BYstresstensor}
 T_{ab}=\frac{1}{8\pi
G_N}\left[Kh_{ab}-K_{ab}+\frac{l}{2}\left(G_{ab}-\frac{6}{l^2}h_{ab}
\right)\right]\,.
\end{equation}
In the expression for $Q_\chi$ by $\Sigma$ we denote an equal time surface at the boundary, i.e. $\Sigma=\Sigma_T\cap\partial\mathcal{M}$ in which $\Sigma_T$ is an equal time $T$ surface of the space-time $\mathcal{M}$. The induced cut-off boundary metric on $\Sigma$ is denoted by $\sigma_{ij}$. The vector $u^\mu$ denotes a future-directed timelike unit normal to the surface
$\Sigma_T$. Specifically let $T^\mu$ be some timelike vector which provides us with a time orientation and which satisfies $T^\mu\partial_\mu T=1$ then we have
\begin{equation}\label{eq:normalsigmat}
 u^\mu=-g^{\mu\nu}\frac{\partial_\nu T}{\sqrt{-g^{TT}}}\,.
\end{equation}

It has been shown in \cite{Gibbons:2004ai,Papadimitriou:2005ii} that $Q_{\partial_T}$ satisfies the quantum statistical relation
\begin{equation}\label{eq:thermodynamicrelation}
Q_{\partial_T}=\beta^{-1}S+\mu_N N+\Omega J+\beta^{-1}I_E\,,
\end{equation}
From this it follows that in terms of $E$ and $\Phi_G$ we have the thermodynamical relation
\begin{equation}\label{eq:thermodynamicpotentials}
E=\beta^{-1}S+\mu_N N+\Omega J+\Phi_G\,.
\end{equation}
The energy $E$ satisfies the first law of thermodynamics
\begin{equation}\label{eq:firstlaw}
dE=\beta^{-1}dS+\mu_N dN+\Omega dJ\,.
\end{equation}
The importance of writing the quantum statistical relation in terms of $E$ and $\Phi_G$ comes from the fact that $E$ and $\Phi_G$ (and not $Q_{\partial_T}$ and $I_E$), which are the only two quantities that we need to compute on the boundary ($N$ and $J$ can be computed as surface integrals over the spatial sections of the horizon), are each independent of the choice of representative of the conformal boundary metric. Hence \eqref{eq:thermodynamicpotentials} is form invariant under PBH transformations that preserve the three boundary Killing vectors $\partial_T$, $\partial_V$ and $\partial_\psi$ \cite{Papadimitriou:2005ii}. 

The thermodynamics of \eqref{eq:secondscaling} and \eqref{eq:blackbrane1} can be obtained by applying the appropriate scaling limits to the thermodynamics of the MMT black hole.

We have tacitly assumed that in the computation of the conserved charges we have been using a frame that does not rotate at infinity. Suppose we were using a frame that does rotate at infinity. Such a frame can be obtained by starting with \eqref{eq:scalinglimit} and performing the following coordinate transformation $T'=T$ and $\psi'=\psi-bT$ where the primed coordinates refer to the rotating frame. The Killing vector $\partial_T$ with respect to which we defined the mass in the frame that is not rotating at the boundary transforms as $\partial_T=\partial_{T'}-b\partial_{\psi'}$ and further $\partial_\psi=\partial_{\psi'}$. Rewriting the Killing vector $X$ that beomes null on the horizon in terms of the primed Killing vectors we find that \eqref{eq:thermodynamicpotentials} becomes
\begin{equation}
 E'=\beta^{-1}S+\mu_N N+\Omega' J+\Phi_G\,,
\end{equation}
where $E'$ is $Q_{\partial_{T'}}-Q_{\partial_{T'}}(\text{$m=0$ solution})$ and where $\Omega'=\Omega+b$. Because the rotating and non-rotating frames have boundary metrics that are diffeomorphic we have that $Q_{\partial_{T'}}(\text{$m=0$ solution})$ is equal to $Q_{\partial_T}(\text{$m=0$ solution})$. The energy $E'$ is related to $E$ via $E'=E+bJ$. The value of the on-shell action is the same in both coordinates systems (also this follows from the fact that the boundary metrics in both frames are diffeomorphic). It can be checked that $E'$ does not satisfy a first law of thermodynamics like \eqref{eq:firstlaw} with $\Omega$ replaced by $\Omega'$. Instead we have
\begin{equation}
 dE'=\beta^{-1}dS+\mu_N dN+\Omega' dJ+Jdb\,.
\end{equation}
Since the grand potential $\Phi_G$ is the same in the rotating and the non-rotating frames we also have
\begin{equation}
 d\Phi_G=S\frac{d\beta}{\beta^2}-Nd\mu_N-Jd\Omega'+Jdb=S\frac{d\beta}{\beta^2}-Nd\mu_N-Jd\Omega\,.
\end{equation}
We conclude that the grand potential in the rotating frame depends on $\Omega$ and not on $\Omega'$. See also \cite{Caldarelli:1999xj,Gibbons:2004ai} for a discussion of the choice of $\Omega$ in the rotating frame.

\subsection{Phase transitions}\label{subsec:thermo}

\begin{figure}
\centering
\psfrag{rH}{$r_H$}
\psfrag{b}{$b$}
\includegraphics{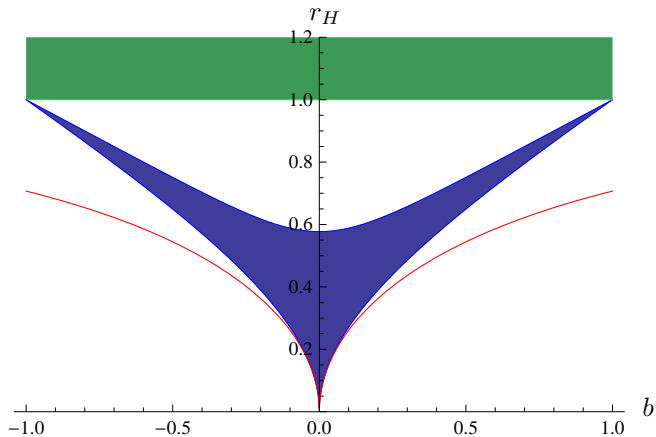}
\caption{Specific heat and grand potential as a function of the black hole parameters $b$ and $r_H$. The green and blue shaded areas are regions where the specific heat is positive and where the grand potential is negative and positive, respectively. The red curve denotes the extremal black holes so that below this curve there are no black hole solutions. The white areas are places where the specific heat is negative and the grand potential is positive.}\label{fig:specificheat}
\end{figure}

The entropy $S$ is given by
\begin{equation}
S=\beta^2\left(\frac{\partial\Phi_G}{\partial\beta}\right)_{\mu_N,\Omega}=\frac{A}{4G}\,,
\end{equation}
where $A$ is the horizon area given in \eqref{eq:horizonarea} and the specific heat is given by
\begin{equation}
-\beta\left(\frac{\partial S}{\partial\beta}\right)_{\mu_N,\Omega}=\frac{\pi^2 L(r_H^2+1)(2r_H^4+b^2(r_H^2-1))(3r_H^4-r_H^2-b^2(r_H^2+1))}{2G_N\Xi_br_H(r_H^2-1)(2r_H^4-b^2(r_H^2+1))}\,.
\end{equation}
In figure \ref{fig:specificheat} we have indicated by green the region where the specific heat is positive and where the grand potential $\Phi_G$ is negative, by blue the region where the specific heat is positive and the grand potential $\Phi_G$ is positive and finally by white the regions where the specific heat is negative and $\Phi_G$ is positive. Further the red curve is the line of extremality at which the specific heat and the temperature vanish. Below this line there are no black hole solutions. Regions where the specific heat is negative are thermodynamically unstable while regions where the specific heat is positive are locally thermodynamically stable.

We will assume that besides the MMT black hole the only other saddle point of the thermodynamical action \eqref{eq:Euclideanaction} that has the same values for the temperature and chemical potentials is plane wave AdS with $iT\sim iT+\beta$, $iV\sim iV+\beta L\mu_N$ and $i\psi\sim i\psi+\beta\Omega$ and with the $V$ coordinate compactified $V\sim V+2\pi L$. The grand potential $\Phi_G$ is the difference between the on-shell action for the Wick rotated MMT black hole and the on-shell action for an equal temperature thermal plane wave AdS$_5$ space-time at finite chemical potentials equal to those of the MMT black hole, which we simply refer to as thermal plane wave AdS$_5$. When $r_H>1$ the grand potential is negative. This means that the black hole is thermodynamically favored over thermal plane wave AdS. Hence a system of pure thermal radiation with a temperature above $\beta^{-1}(r_H=1)$ and chemical potentials $\mu_N<0$ and $\Omega$ will form a MMT black hole. On the other hand, since the grand potential is positive for $r_H<1$, thermal plane wave AdS with chemical potentials $\mu_N>0$ and $\Omega$ will be thermodynamically favored. Since in the white regions of figure \ref{fig:specificheat} we have a negative specific heat and $\Phi_G>0$ we expect that they correspond to a phase of pure radiation. 

We will argue that the black holes in the blue region are classically unstable. In subsection \ref{subsec:geometricproperties} we saw that the Killing vector $X^\mu$ that becomes null at the horizon is everywhere timelike outside the horizon when $r_H>1$. In \cite{Hawking:1999dp} it has been proven that in this case there is no superradiant scattering of waves incident on the black hole. When $r_H<1$ we observed that $X^\mu$ becomes spacelike in a large region of the space-time that lies strictly outside the horizon and that extends to the boundary. It follows that there does not exist any Killing vector field which remains timelike everywhere outside the horizon. Subsequently, in this case superradiant scattering does occur and since the space-time is asymptotically AdS, superradiant modes are reflected back into the bulk either due to the gravitational potential well at infinity or due to boundary conditions. This leads to a classical instability of black holes with $r_H<1$.

The blue shaded region of figure \ref{fig:specificheat} seems to form a different phase than the surrounding white shaded regions. The specific heat is divergent as one goes from the white regions, that are in between the red extremality curve and the blue region, to the blue region. The black holes in the blue region are thermodynamically disfavored with respect to plane wave AdS and they are classically unstable but they have positive specific heat so depending on the time scales involved we may refer to this as a metastable phase. 

The thermodynamic behavior of the MMT black hole is reminiscent but slightly different from the Hawking--Page phase transition \cite{Hawking:1982dh}. Since we have proven that the thermodynamics is TsT invariant and since both the saddle points, thermal plane wave AdS$_5$ and the MMT black hole have a TsT transformed version and conversely since any saddle point after TsT which has the same Killing vectors has a description before TsT as an AAdS$_5$ metric, we should observe the same behavior with thermal plane wave AdS$_5$ replaced by global Sch$_5$ and the MMT black hole replaced by its TsT transformed version.

\section{Discussion}

The way we employed the TsT transformation is sort of in reverse. We use it to parametrize the metric of an ASch$_5$ space-time in terms of a metric that is defined on a space-time (in this case pure AAdS$_5$) that, unlike the ASch$_5$ space-time, has a conformal boundary. A central role in the TsT transformation is played by the vector $N=\partial_V$. We have identified a conformal class of Schr\"odinger boundary metrics that are the TsT transform of those AdS boundaries that possess a null Killing vector. The TsT transformation then guarantees that all AAdS$_5$ properties that `commute' with $N$ carry over to the ASch$_5$ case. For Schr\"odinger boundary metrics that can be viewed as the TsT transform of an AdS boundary metric possessing a null Killing vector we constructed local counterterms. These counterterms are invariant under boundary diffeomorphisms as well as PBH transformations that respect the property that $N$ is a boundary null Killing vector (possibly up to some nonrelativistic conformal anomaly). These transformations together form the local symmetries of the boundary theory.

Another attractive feature of these Schr\"odinger boundaries is that in the pure Sch$_5$ case the $V$ coordinate can be thought of as parametrizing the light like lines \cite{Blau:2010fh} which in turn form the boundaries of the light cones of the space-time. The Galilean-like causal structure is then naturally inherited by boundaries that have $V$ as a tangential coordinate. 

It is tempting to state that the near boundary region of any ASch$_5$ space-time can be thought of as the TsT transformation of the near boundary region of an AAdS$_5$ space-time and it would be interesting to see if the TsT relation between ASch$_5$ and AAdS$_5$ solutions can be relaxed in the interior of the space-time. When relaxing some of the properties of the TsT transformation it may no longer be a solution generating technique, but it could provide the starting point for a suitable Ansatz for solving the equations of motion \eqref{eq:EinsteineqTsT} to \eqref{eq:scalarTsT}.

One possibility is to relax the condition that $N$ is a global Killing vector of the space-time to the condition that it is only an asymptotic Killing vector. Such solutions break particle number in the bulk of the space-time. These space-times can be physically interesting for the following reason. The Schr\"odinger invariant system of fermions at unitarity possesses a ground state that spontaneously breaks particle number (see \cite{Bloch:2008zzb} for a nice review of cold atom systems and see \cite{Balasubramanian:2010uw} for a discussion in the context of holography) and so it would be interesting to study geometries that break particle number away from the boundary. Another (perhaps related to the presence of a Killing vector) special feature of the TsT transformed AAdS$_5$ metrics is the relation between the mass term $A^2$ and the potential $V(\Phi)$ which in the bulk action cancel each other so as to leave a pure cosmological constant. We could try to relax this condition by demanding that this happens only near the boundary but not in the interior of the space-time.

We further showed that the on-shell action as well as the thermodynamic properties of black hole space-times that relate to the horizon such as temperature, chemical potentials and entropy are all TsT invariant. We used this to show that there is a Hawking--Page type phase transition between global Schr\"odinger space-time at finite temperature and chemical potentials and the TsT transformed MMT black hole. It would be interesting to find a dual field theory interpretation (much like as was done for the usual Hawking--Page phase transition in \cite{Witten:1998zw}) for this phase transition and in particular to explain what the role of the harmonic trapping potential is. 

Hand in hand with constructing more general ASch$_5$ solutions another interesting next step would be the formulation of a well-posed variational problem and the related question as to what the right notion of a boundary stress energy tensor should be (see \cite{Adams:2008wt,Ross:2009ar,Guica:2010sw} for some proposals) and to see what the associated algebra of conserved charges is. We hope to report on some of these questions in the future.

\section*{Acknowledgments}
We wish to thank Emiliano Imeroni for collaboration in initial stages of this research and for many useful discussions. Further we wish to thank Bom Soo Kim, Balt van Rees, Kostas Skenderis, Marika Taylor and  Daiske Yamada for useful discussions. Finally we express our gratitude to Matthias Blau for many useful discussions and careful reading of this manuscipt. This work was supported in part by the Swiss National Science Foundation and the ``Innovations- und Kooperationsprojekt C-13'' of the Schweizerische Universit\"atskonferenz SUK/CRUS.

\appendix

\section{TsT transformations}\label{app:TsT}

We review the details of a TsT transformation which has the direction for the T-duality in the 5-sphere and the direction for the shift in the pure AAdS$_5$ space-time. We closely follow \cite{Maldacena:2008wh,Imeroni:2009cs}.

Consider the following 10-dimensional Einstein frame metric
\begin{equation}
 ds^2_E=g_{\mu\nu}dx^\mu dx^\nu+ds^2_{S^5}\,,
\end{equation}
where $ds^2_{S^5}$ is the metric on a unit radius 5-sphere and where $g_{\mu\nu}dx^\mu dx^\nu$ is any solution to the 5-dimenional Einstein equations
\begin{equation}\label{eq:EinsteineqsFR}
 G_{\mu\nu}-6g_{\mu\nu}=0\,.
\end{equation}
The above 10-dimensional metric is a solution of type IIB supergravity (Freund--Rubin compactification) if we also switch on a 5-form flux given by
\begin{equation}
 F_5=(1+\star)G_5\,,
\end{equation}
in which $G_5=4\frac{1}{5!}\epsilon_{\mu_1\ldots\mu_5}dx^{\mu_1}\wedge\cdots\wedge dx^{\mu_5}$ with $\epsilon_{\mu_1\ldots\mu_5}=\sqrt{-g}e_{\mu_1\ldots\mu_5}$ denoting the volume form on $g_{\mu\nu}dx^\mu dx^\nu$, where $e_{\mu_1\ldots\mu_5}$ is the Levi-Civit\`a symbol. We use a normalization of the 5-form such that the 10-dimensional Einstein equations read
\begin{equation}
 G_{MN}=\frac{1}{96}F_{MPQRS}F_N{}^{PQRS}+\ldots\,.
\end{equation}
All other type IIB fields are zero.

The 5-sphere metric can be written, using a Hopf fibration, as
\begin{equation}
 ds^2_{S^5}=d\eta^2+ds^2_{\mathbb{CP}^2}\,,
\end{equation}
in which $\eta$ is such that $d\eta/2$ equals the K\"ahler 2-form of the base space $\mathbb{CP}^2$. Writing $\eta=d\xi+P$ the coordinate $\xi$ parameterizes a Killing direction, i.e. $\partial_\xi$ is a Killing vector (the Reeb vector).

Let $\partial_V$ denote a Killing vector of the metric $g_{\mu\nu}dx^\mu dx^\nu$. Performing a T-duality along $\xi$ with the T-dual circle being parametrized by $\tilde\xi$, subsequently shifting $V\rightarrow V+\gamma\tilde\xi$ and performing a second T-duality along $\tilde\xi$ leads to the following solution of type IIB supergravity which in Einstein frame reads \cite{Maldacena:2008wh,Imeroni:2009cs}
\begin{eqnarray}
 ds^2_E & = & e^{-\Phi/2}\left(g_{\mu\nu}-e^{-2\Phi}A_\mu A_\nu\right)dx^\mu dx^\nu+e^{-\Phi/2}ds^2_{\mathbb{CP}^2}+e^{3\Phi/2}\eta^2\,,\\
B & = & A\wedge\eta\,,\\
F_5 & = & 4\sqrt{-g}\frac{1}{5!}e_{\mu_1\ldots\mu_5}dx^{\mu_1}\wedge\cdots\wedge dx^{\mu_5}+4\eta\wedge\text{Vol}(\mathbb{CP}^2)\,,
\end{eqnarray}
where $B$ denotes the NS-NS 2-form and where the 5-dimensional vector $A$ and scalar $\Phi$ (dilaton) are given by
\begin{eqnarray}
 A&=&\gamma e^{2\Phi}g_{V\mu}dx^\mu\,,\label{eq:TsTvector}\\
e^{-2\Phi} & = & 1+\gamma^2g_{VV}\,.\label{eq:TsTscalar}
\end{eqnarray}
The remaining IIB fields (RR potentials) are zero.

The idea is now to reduce the Einstein frame type IIB action over the squashed 5-sphere down to five dimensions. In doing so we obtain the 5-dimensional action and the 5-dimensional Einstein frame metric which we are interested in. Using the results of \cite{Maldacena:2008wh} this procedure gives rise to the following 5-dimensional Einstein frame metric
\begin{equation}
 ds^2 =\bar g_{\mu\nu}dx^\mu dx^\nu=e^{-2\Phi/3}\left(g_{\mu\nu}-e^{-2\Phi}A_\mu A_\nu\right)dx^\mu dx^\nu\,.\label{eq:TsTmetric}
 \end{equation}
This metric together with the vector \eqref{eq:TsTvector} and scalar field \eqref{eq:TsTscalar} solve the equations of motion coming from the following action
\begin{eqnarray}
 I_{\text{bulk}}+I_{\text{GH}} & = & \frac{1}{16\pi G_N}\int_{M}d^5x\sqrt{-\bar g}\left(\bar R-\frac{4}{3}\partial_\mu\Phi\partial^\mu\Phi-V(\Phi)-\right.\nonumber\\
&&\left.\frac{1}{4}e^{-\tfrac{8}{3}\Phi}F_{\mu\nu}F^{\mu\nu}-4A_\mu A^\mu\right)+\frac{1}{8\pi G_N}\int_{\partial\mathcal{M}}d^4\xi\sqrt{-\bar h}\bar K\,,\nonumber\\
&&\label{eq:5Daction}
\end{eqnarray}
where the potential $V$ is given by
\begin{equation}
 V(\Phi)=4e^{\tfrac{2}{3}\Phi}\left(e^{2\Phi}-4\right)\,.
\end{equation}
This potential has an absolute minimum at $\Phi=0$ where it assumes the value $-12$. For large negative values of $\Phi$ it asymptotes to zero and for large positive values it blows up.

The bulk equations of motion that follow from this action are
\begin{align}
 & \bar G_{\mu\nu} = \frac{4}{3}\left(\partial_\mu\Phi\partial_\nu\Phi-\frac{1}{2}\partial_\rho\Phi\partial^\rho\Phi \bar g_{\mu\nu}\right)-\frac{1}{2}V(\Phi)\bar g_{\mu\nu}\label{eq:EinsteineqTsT}\\
&+\frac{1}{2}e^{-\tfrac{8}{3}\Phi}\left(F_{\rho\mu}F^\rho{}_\nu-\frac{1}{4}F_{\rho\sigma}F^{\rho\sigma}\bar g_{\mu\nu}\right)+4A_\mu A_\nu-2A_\rho A^\rho \bar g_{\mu\nu}\,,\nonumber\\
&\frac{1}{\sqrt{-\bar g}}\partial_\mu\left(\sqrt{-\bar g}e^{-\tfrac{8}{3}\Phi}F^{\mu\nu}\right) = 8A^\nu\,,\label{eq:massivevectoreqTsT}\\
&\frac{1}{\sqrt{-\bar g}}\partial_\mu\left(\sqrt{-\bar g}\partial^\mu\Phi\right) = \frac{3}{8}V'(\Phi)-\frac{1}{4}e^{-\tfrac{8}{3}\Phi}F_{\mu\nu}F^{\mu\nu}\,.\label{eq:scalarTsT}
\end{align}
From the equation for $A^\mu$ one derives that
\begin{equation}
 \partial_\mu\left(\sqrt{-\bar g}A^\mu\right)=0\,.
\end{equation}

The reduction over the squashed 5-sphere is consistent and forms a special case of dimensional reductions over squashed Sasaki--Einstein manifolds \cite{Cassani:2010uw,Gauntlett:2010vu,Liu:2010sa,Skenderis:2010vz,Bah:2010cu}.

We discuss the isometries that are preserved by the TsT transformation. Suppose that $K$ is a Killing vector of the metric $g_{\mu\nu}$. The Lie derivative of $\bar g_{\mu\nu}$ along $K$ is given by
\begin{eqnarray}
\mathcal{L}_K \bar g_{\mu\nu} & = & \gamma e^{-2\Phi/3}\left(-\frac{4}{3}e^{-2\Phi}A_\mu A_\nu A_\rho+\right.\nonumber\\
&&\left.\left(g_{\mu\rho}A_\nu+g_{\nu\rho}A_\mu-\frac{2}{3}g_{\mu\nu}A_\rho\right)\right)[\partial_V,K]^\rho\,.\label{eq:KillingTsT}
\end{eqnarray}
Hence, only those Killing vectors $K$ that commute with $\partial_V$ are also Killing vectors of $\bar g_{\mu\nu}$. Conversely, given a Killing vector of the metric $\bar g_{\mu\nu}$ then it must also be a Killing vector of $g_{\mu\nu}$. To see this note that
\begin{eqnarray}
 0=\bar g_{\rho\nu}\bar\nabla_\mu K^\rho+\bar g_{\rho\mu}\bar\nabla_\nu K^\rho & = & g_{\rho\nu}\nabla_\mu K^\rho+g_{\rho\mu}\nabla_\nu K^\rho\nonumber\\
&&+\text{terms proportional to $\gamma$}\,.
\end{eqnarray}
Since the first term on the right hand side only involves the AdS metric, which cannot depend on $\gamma$, this term has to vanish by itself. We conclude that the commutant of the Killing vector $\partial_V$ used in the TsT transformation forms the complete isometry algebra of the metric $\bar g_{\mu\nu}$.

We collect some TsT transformation formulas. In the bulk we have
\begin{eqnarray}
\bar g^{\mu\nu} & = & e^{2\Phi/3}g^{\mu\nu}+e^{-10\Phi/3}g^{\mu\rho}g^{\nu\sigma}A_\rho A_\sigma\,,\label{eq:bulkTsT1}\\
\sqrt{-\bar g} & = & e^{-2\Phi/3}\sqrt{-g}\,,\label{eq:detgTsT}\\
\bar g^{\rho\sigma}A_\sigma F_{\rho\mu} & = & 2e^{8\Phi/3}\partial_\mu\Phi\,,\label{eq:TsTidentity1}\\
\bar g^{\mu\sigma}\bar g^{\rho\tau}F_{\mu\rho}F_{\sigma\tau} & = & e^{4\Phi/3}F_{\mu\rho}F_{\sigma\tau}g^{\mu\sigma}g^{\rho\tau}+8e^{16\Phi/3}g^{\rho\tau}\partial_\rho\Phi\partial_\tau\Phi\,,\\
\bar g^{\mu\nu}\partial_\mu\Phi\partial_\nu\Phi & = & e^{2\Phi/3}g^{\mu\nu}\partial_\mu\Phi\partial_\nu\Phi\,,\\
\bar g^{\mu\nu}A_\mu A_\nu & = & e^{2\Phi/3}(1-e^{2\Phi})\,,\label{eq:bulkTsT6}\\
 \bar R&=&e^{2\Phi/3}\left( R  + \left(\frac{4}{3} + 2 e^{2\Phi}
\right)g^{\mu\nu}\partial_\mu \Phi \partial_\nu \Phi\right. \nonumber\\
&& \left.+\frac{1}{4}e^{-2\Phi}F_{\mu\nu}F_{\rho\sigma}g^{\mu\rho}g^{\nu\sigma} +\frac{2}{3}\square\Phi\right)\,.\label{eq:bulkTsT7}
\end{eqnarray}
and on the boundary we have
\begin{eqnarray}
 \bar h^{ab} & = & e^{2\Phi/3}h^{ab}+e^{-10\Phi/3}h^{ac}h^{bd}A_cA_d\,,\label{eq:bdryTsT1}\\
 \sqrt{-\bar h} & = & e^{-\Phi/3}\sqrt{-h}\,,\label{eq:detboundaryTsT}\\
\bar K & = & e^{\Phi/3}\left(K-\frac{1}{3}n^\mu\partial_\mu\Phi\right)\,,\label{eq:bdryTsT3}\\
\bar R_{(\bar h)} & = & e^{2\Phi/3}\left[R_{(h)}+\frac{1}{4}e^{-2\Phi}F_{ab}F_{cd}h^{ac}h^{bd}\right.\nonumber\\
&&\left.+\left(\frac{4}{3}+2e^{2\Phi}\right)h^{ab}\partial_a\Phi\partial_b\Phi\right]\,.\label{eq:TsTRh}
\end{eqnarray}
In deriving equations \eqref{eq:bulkTsT1} to \eqref{eq:bulkTsT7} we used \eqref{eq:TsTvector} to \eqref{eq:TsTmetric} as well as the fact that $\partial_V$ is a bulk Killing vector. In deriving \eqref{eq:bdryTsT1} to \eqref{eq:TsTRh} we used that the boundary metric is $\bar h_{ab}=e^{-2\Phi/3}h_{ab}-e^{-8\Phi/3}A_aA_b$ with $A_a=\gamma e^{2\Phi}h_{Va}$ and $e^{-2\Phi}=1+\gamma^2 h_{VV}$ and assumed that $\partial_V$ is a Killing vector of $h_{ab}$. These formulas are true for any TsT transformation and not just the ones that give rise to asymptotically Schr\"odinger space-times.

In proving \eqref{eq:detgTsT} we used that $\text{det}(1-C)=1-\text{Tr}C$ for a matrix $C$ that is the product of two vectors. In establishing \eqref{eq:detboundaryTsT} it is important that $V$ is tangential to the boundary. In that case we can say the following. Suppose that we would compute the induced metric on a co-dimension $n$ surface that has $V$ as one of its tangential directions. The induced metric $\bar h_{(n)ij}$ is
\begin{equation}
\bar h_{(n)ij}=\delta^\mu_i\delta^\nu_j\bar g_{\mu\nu}\,,
\end{equation}
where $i=(V,I)$ with $I=1,\ldots,4-n$. Then we find for the determinant $\text{det}(\bar h_{(n)ij})$ of the induced metric
\begin{equation}
\text{det}(\bar h_{(n)ij}) = e^{2(n-2)\Phi/3}\text{det}(h_{(n)ij})\,,\label{eq:detinducedTsT}
\end{equation}
where $h_{(n)ij}=\delta^\mu_i\delta^\nu_jg_{\mu\nu}$.

\end{document}